\documentclass[letterpaper]{jpconf}
\usepackage{graphicx}
\begin{document}
\title{A Survey of Multiplicity Fluctuations in PHENIX}

\author{J.T. Mitchell and the PHENIX Collaboration}

\address{Brookhaven National Laboratory, P.O. Box 5000, Building 510C, Upton, NY 11973-5000}

\ead{mitchell@bnl.gov}

\begin{abstract}
The PHENIX Experiment at the Relativistic Heavy Ion Collider has made measurements of event-by-event fluctuations in the charged particle multiplicity as a function of collision energy, centrality, collision species, and transverse momentum in heavy ion collisions. The results of these measurements will be reviewed and discussed.
\end{abstract}.

\section{Introduction}

Interest in the topic of charged particle multiplicity fluctuations has been recently revived by the observation of non-monotonic behavior in the scaled variance as a function of system size at SPS energies \cite{na49MF}.  The scaled variance is defined as $var(N)/<N>$, where $var(N)$ represents the variance of the multiplicity distribution in a given centrality bin, and $<N>$ is the mean of the distribution. For reference, the scaled variance of a Poisson distribution is 1.0, independent of N. PHENIX has studied the behavior of charged particle multiplicity fluctuations as a function of centrality and transverse momentum in $\sqrt{s_{NN}}$ = 62 GeV and 200 GeV Au+Au and Cu+Cu collisions in order to investigate if this behavior persists at RHIC energies.

Details about the PHENIX experimental configuration can be found elsewhere \cite{phenixNIM}. All of the measurements described here utilized the PHENIX central arm detectors. The maximum PHENIX acceptance of $|\eta|<0.35$ in pseudorapidity and $180^{o}$ in azimuthal angle is considered small for event-by-event measurements. However, the event-by-event multiplicities are high enough in RHIC heavy ion collisions that PHENIX has a competitive sensitivity for the detection of many fluctuation signals. For example, a detailed examination of the PHENIX sensitivity to temperature fluctuations derived from the measurement of mean $p_{T}$ fluctuations is described in \cite{ppg005}.

\begin{figure}[h]
\begin{minipage}{18pc}
\includegraphics[width=18pc]{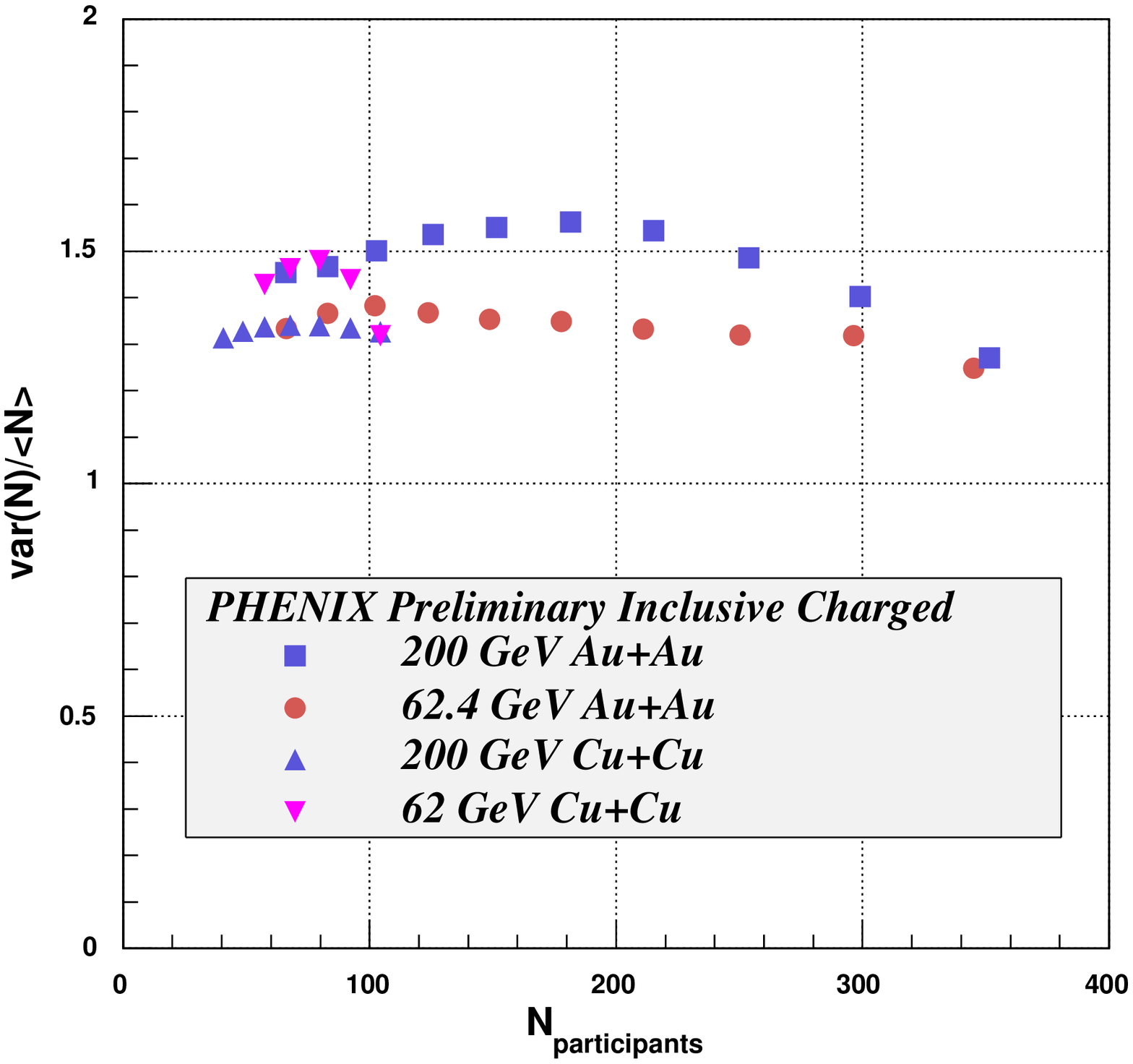}
\caption{\label{fig:varVsCentAll}Inclusive charged particle multiplicity fluctuations in terms of the scaled variance as a function of centrality for $\sqrt{s_{NN}}$ = 62 and 200 GeV Au+Au and Cu+Cu collisions. The error bars include statistical and systematic errors.}
\end{minipage}\hspace{2pc}%
\begin{minipage}{18pc}
\includegraphics[width=18pc]{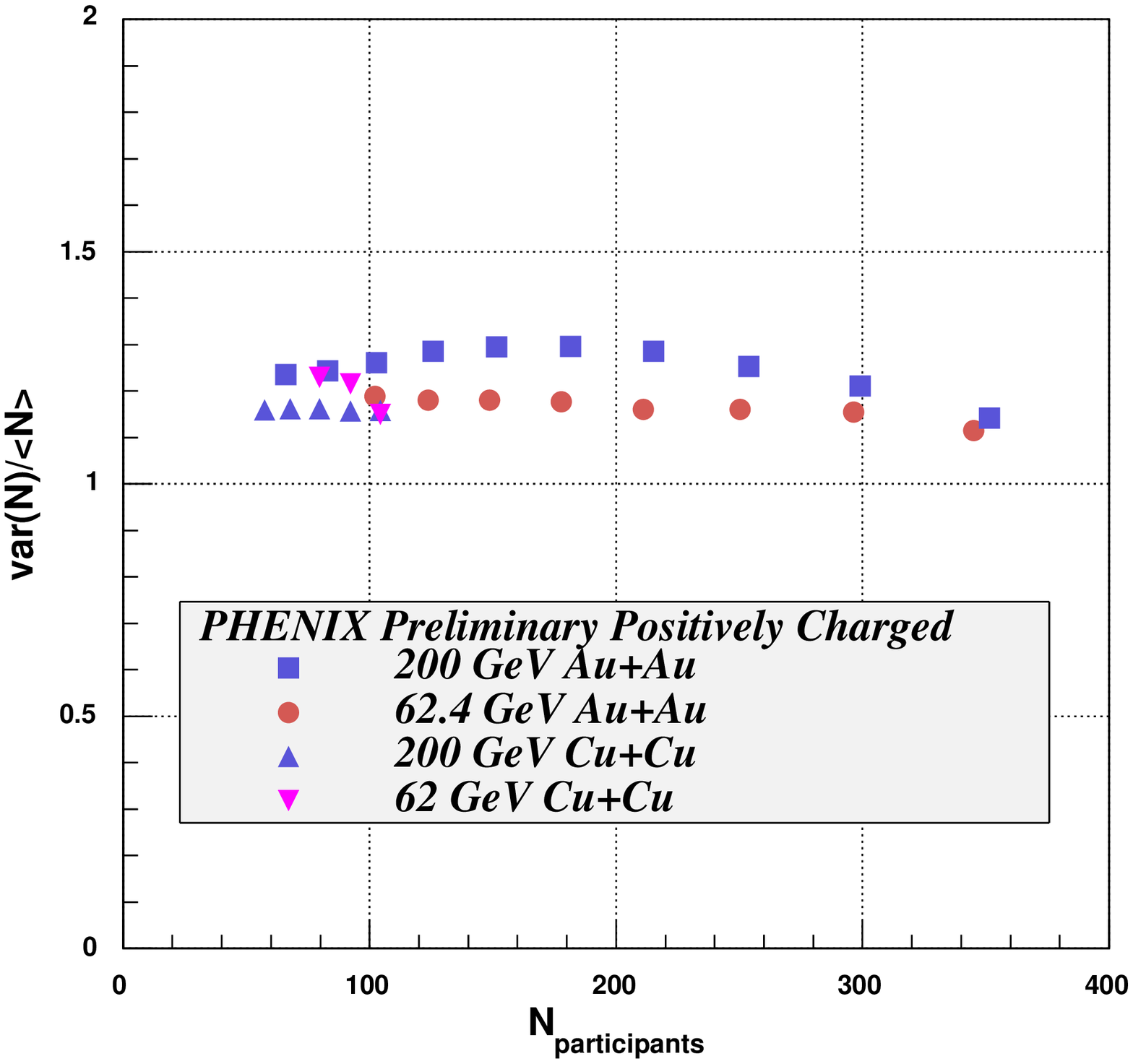}
\caption{\label{fig:varVsCentPos}Positively charged particle multiplicity fluctuations in terms of the scaled variance as a function of centrality for $\sqrt{s_{NN}}$ = 62 and 200 GeV Au+Au and Cu+Cu collisions. The error bars include statistical and systematic errors.}
\end{minipage}\hspace{2pc}%
\begin{minipage}{18pc}
\includegraphics[width=18pc]{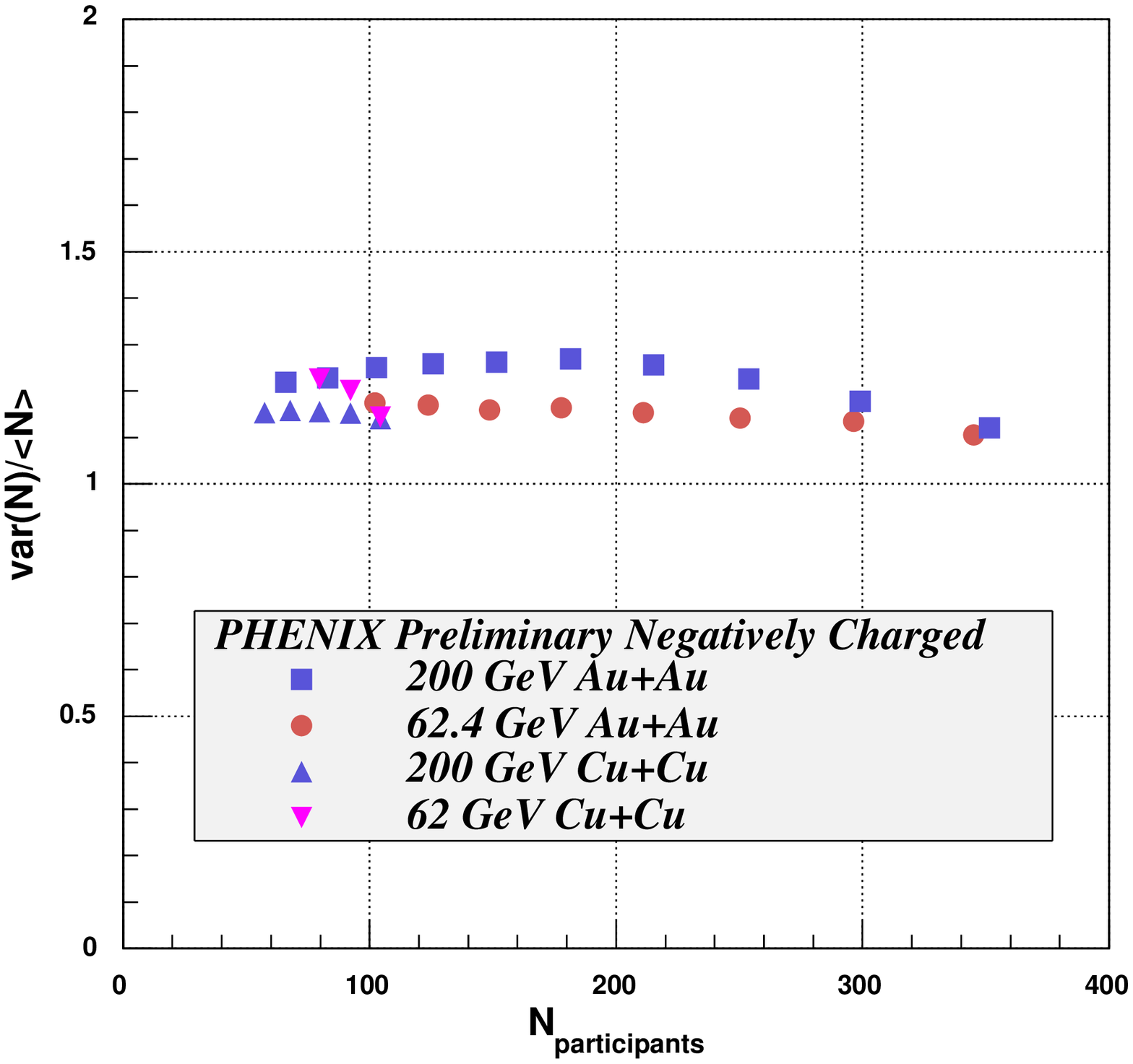}
\caption{\label{fig:varVsCentNeg}Negatively charged particle multiplicity fluctuations in terms of the scaled variance as a function of centrality for $\sqrt{s_{NN}}$ = 62 and 200 GeV Au+Au and Cu+Cu collisions. The error bars include statistical and systematic errors.}
\end{minipage} 
\end{figure}

Charged particle multiplicity fluctuations in terms of the scaled variance as a function of centrality are shown in Fig. \ref{fig:varVsCentAll} for inclusive charged particles, Fig. \ref{fig:varVsCentPos} for positive particles, and Fig. \ref{fig:varVsCentNeg} for negative particles. Contributions from fluctuations due to variations of the impact parameter with in a centrality bin, or geometry fluctuations, are included in these measurements. Fluctuations from 62 GeV Au+Au collisions are more Poissonian than in 200 GeV collisions. There is also a peak structure in 200 GeV Au+Au collisions that is not present at 62 GeV. HIJING studies indicate that this structure is likely attributable to the increased influence of hard processes at 200 GeV.  However, the trend is reversed in Cu+Cu collisions, whereby the 62 GeV fluctuations are less Poissonian than the 200 GeV fluctuations. The centrality-dependence of the fluctuations in 62 GeV Cu+Cu collisions behave qualitatively like measurements reported by NA49 at SPS energies. The qualitative features of the data at both of the RHIC energies are the same for inclusive charge, positively charged, and negatively charged particle distributions. Charge selection at RHIC energies appears only to divide the value of $(var(N)/<N>)-1$ roughly by a factor of two.

\begin{figure}[h]
\begin{minipage}{18pc}
\includegraphics[width=18pc]{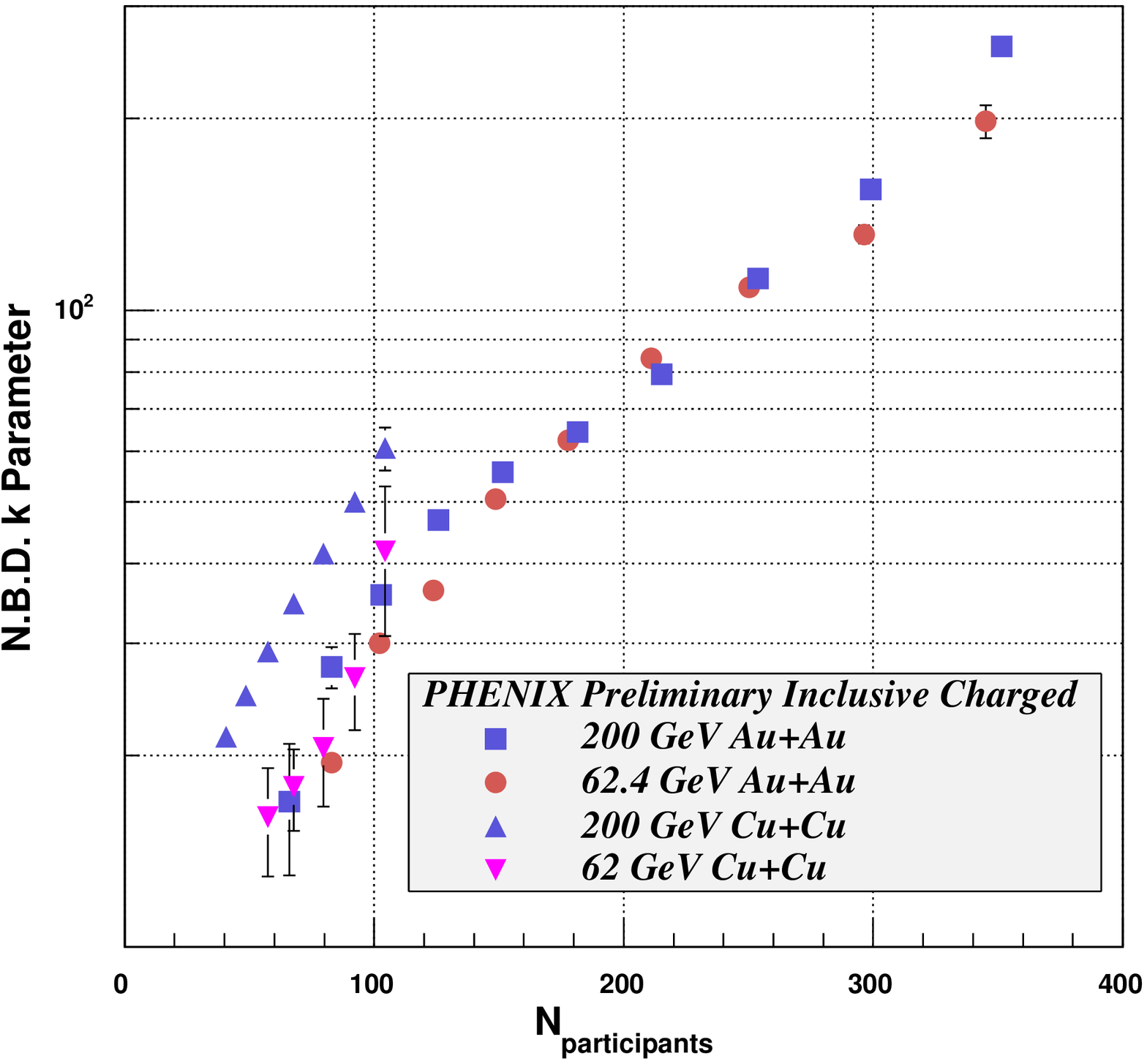}
\caption{\label{fig:kVsCentAll}Inclusive charged particle multiplicity fluctuations in terms of the k parameter from a negative binomial distribution fit to the data as a function of centrality for $\sqrt{s_{NN}}$ = 62 and 200 GeV Au+Au and Cu+Cu collisions.}
\end{minipage}\hspace{2pc}%
\begin{minipage}{18pc}
\includegraphics[width=18pc]{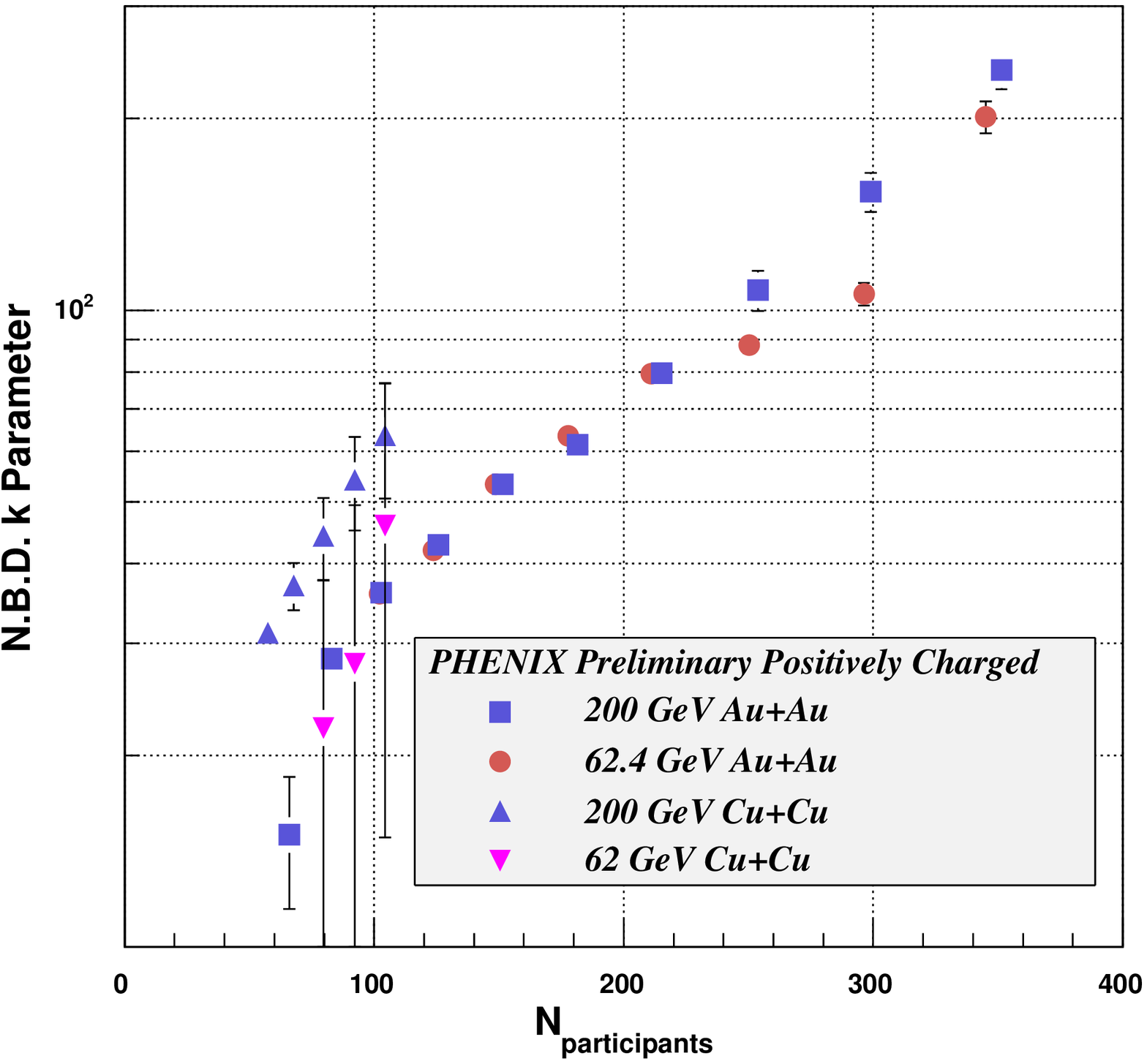}
\caption{\label{fig:kVsCentPos}Positively charged particle multiplicity fluctuations in terms of the k parameter from a negative binomial distribution fit to the data as a function of centrality for $\sqrt{s_{NN}}$ = 62 and 200 GeV Au+Au and Cu+Cu collisions.}
\end{minipage}\hspace{2pc}%
\begin{minipage}{18pc}
\includegraphics[width=18pc]{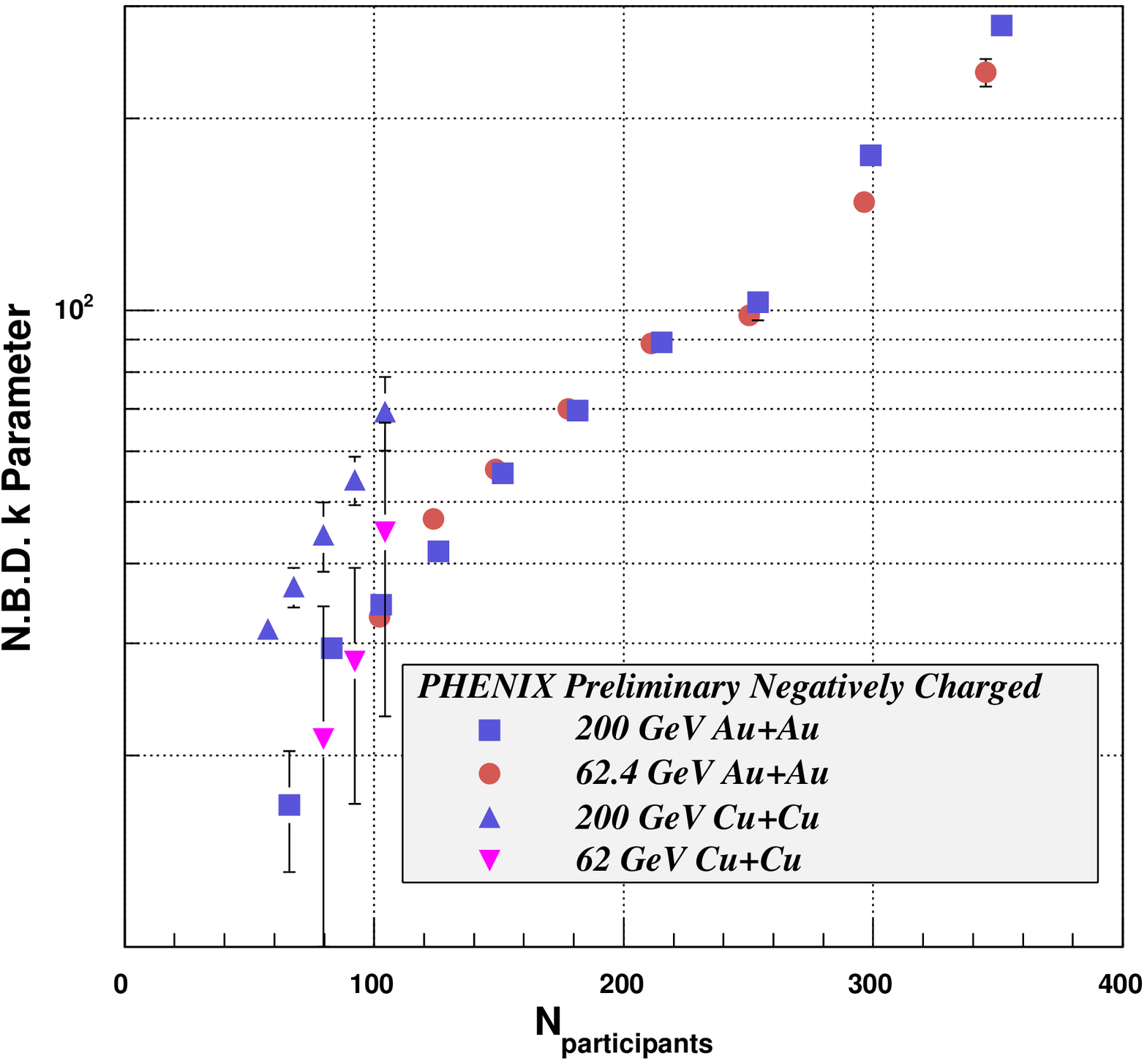}
\caption{\label{fig:kVsCentNeg}Negatively charged particle multiplicity fluctuations in terms of the k parameter from a negative binomial distribution fit to the data as a function of centrality for $\sqrt{s_{NN}}$ = 62 and 200 GeV Au+Au and Cu+Cu collisions.}
\end{minipage} 
\end{figure}

It can be demonstrated that charged particle multiplicity fluctuation distributions in elementary and heavy ion collisions are well described by negative binomial distributions (NBD) \cite{e802MF}. The NBD of an integer $m$ is defined by
\begin{equation}
P(m) = \frac{(m+k-1)!}{m!(k-1)!} \frac{(\mu/k)^{m}}{(1+\mu/k)^{m+k}}
\end{equation}
where $P(m)$ is normalized for $0\leq m \leq \infty$, $\mu\equiv<m>$. The NBD contains an additional parameter, $k$, when compared to a Poisson distribution. The NBD becomes a Poisson distribution in the limit $k\rightarrow\infty$ - the larger the value of $k$, the more Poisson-like is the distribution. The variance and the mean of the NBD is related to $k$ by $1/k = \sigma^{2}/\mu^{2} - 1/\mu$, where $\mu$ is the mean and $\sigma$ is the standard deviation of the distribution. The PHENIX multiplicity distributions are well described by NBD fits for all centralities. The extracted values of the $k$ parameter are shown in Fig. \ref{fig:kVsCentAll}-\ref{fig:kVsCentNeg} as a function of centrality for inclusive charged, positively charged, and negatively charged particles. Except for the most central collisions, the values of $k$ for the 62 and 200 GeV Au+Au data are consistent with each other for all charge selections. There is also little difference in $k$ for the different charge selections.

\begin{figure}[h]
\begin{minipage}{18pc}
\includegraphics[width=18pc]{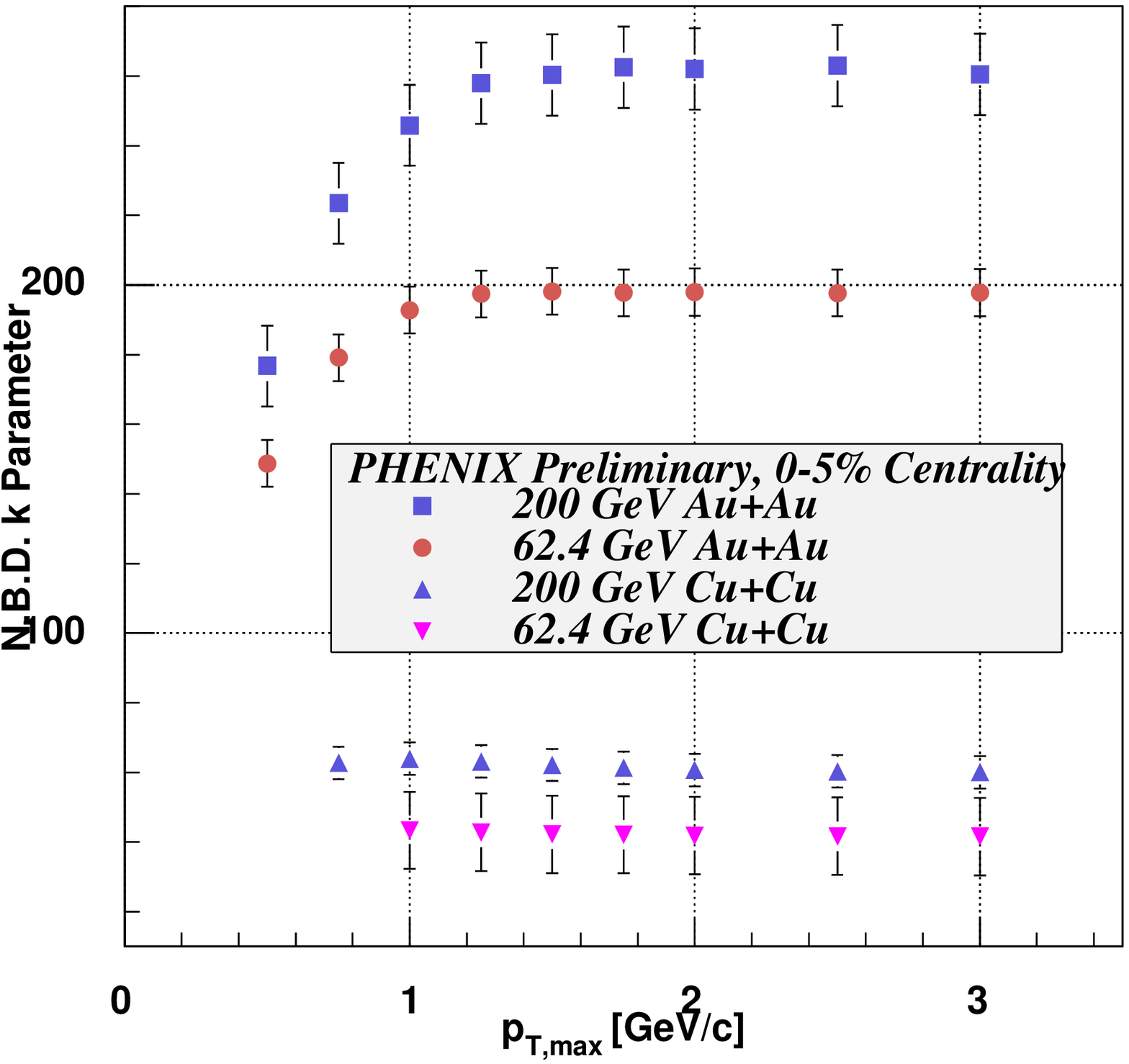}
\caption{\label{fig:kVsPtC0}Inclusive charged particle multiplicity fluctuations in terms of the k parameter from a negative binomial distribution fit to the data as a function of $p_{T}$ over the range 200 MeV/c $<p_{T}<p_{T,max}$ for $\sqrt{s_{NN}}$ = 62 and 200 GeV Au+Au and Cu+Cu 0-5\% central collisions.}
\end{minipage}\hspace{2pc}%
\begin{minipage}{18pc}
\includegraphics[width=18pc]{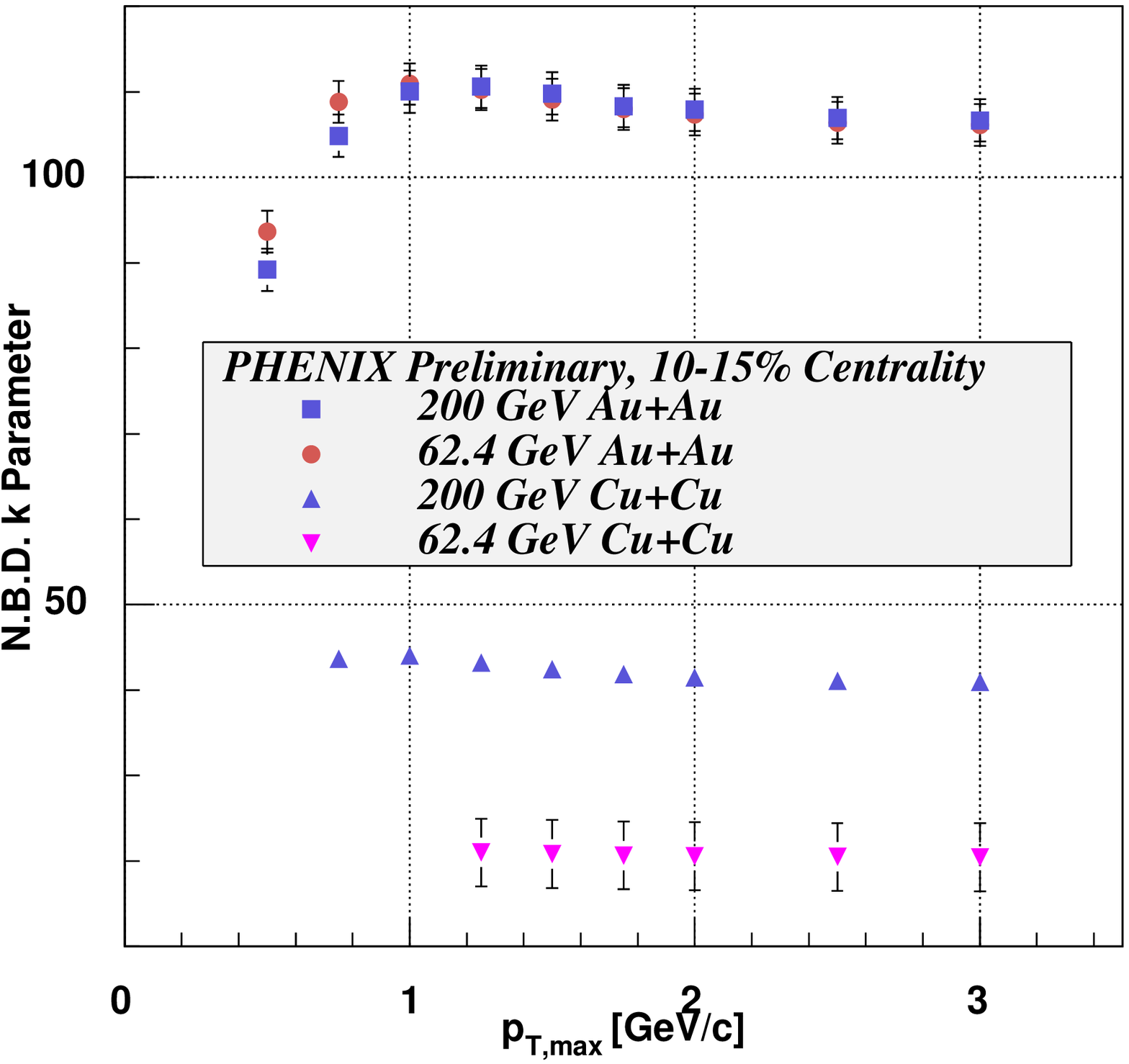}
\caption{\label{fig:kVsPtC2}Inclusive charged particle multiplicity fluctuations in terms of the k parameter from a negative binomial distribution fit to the data as a function of $p_{T}$ over the range 200 MeV/c $<p_{T}<p_{T,max}$ for $\sqrt{s_{NN}}$ = 62 and 200 GeV Au+Au and Cu+Cu 10-15\% central collisions.}
\end{minipage}\hspace{2pc}%
\begin{minipage}{18pc}
\includegraphics[width=18pc]{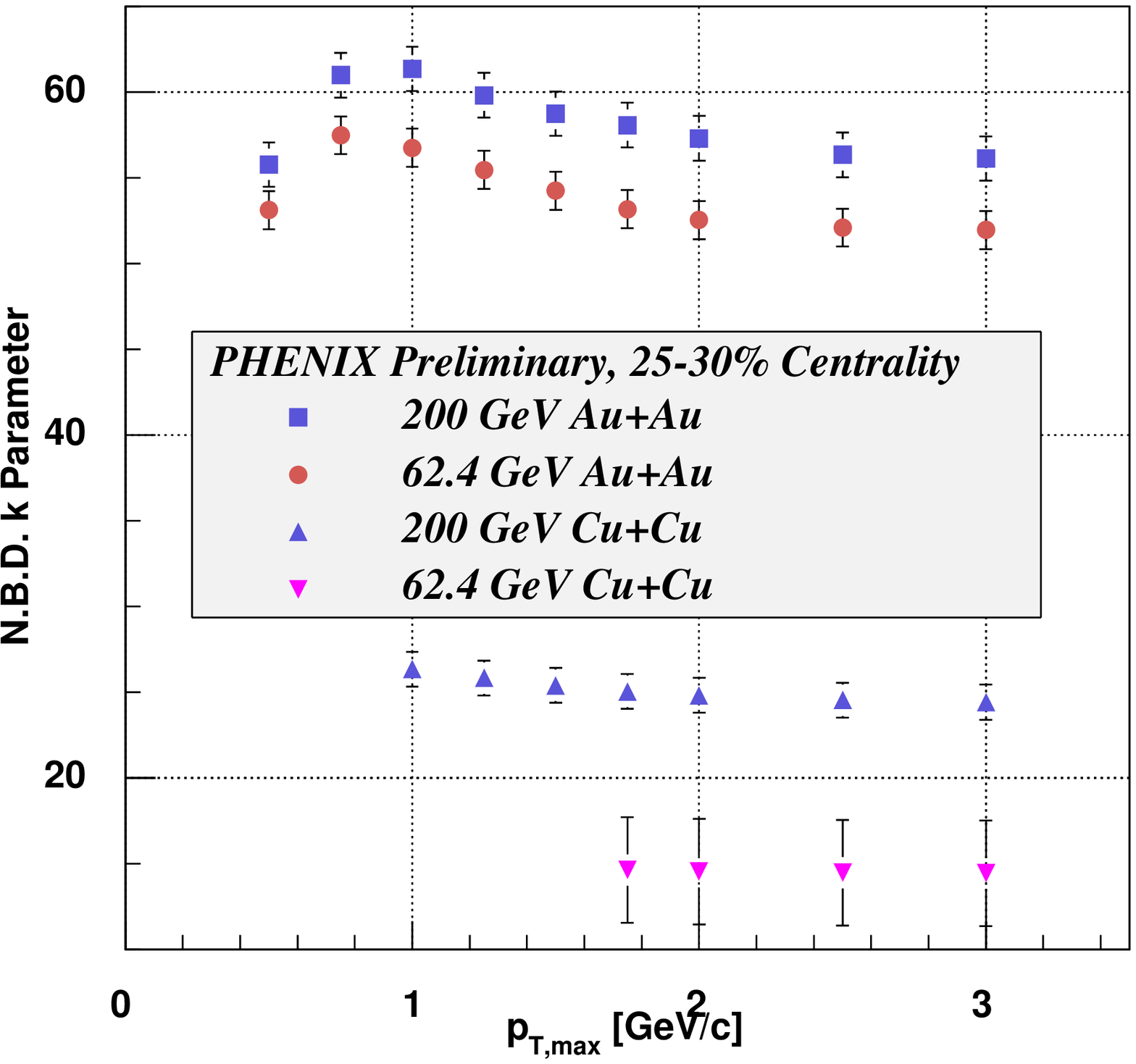}
\caption{\label{fig:kVsPtC5}Inclusive charged particle multiplicity fluctuations in terms of the k parameter from a negative binomial distribution fit to the data as a function of $p_{T}$ over the range 200 MeV/c $<p_{T}<p_{T,max}$ for $\sqrt{s_{NN}}$ = 62 and 200 GeV Au+Au and Cu+Cu 25-30\% central collisions.}
\end{minipage} 
\end{figure}

In order to better understand the various sources of the multiplicity fluctuations, the fluctuations have been measured as a function of $p_T$ range, $200~MeV/c<p_{T}<p_{T,max}$. The results in terms of the NBD $k$ parameter are shown for three centralities in Fig. \ref{fig:kVsPtC0}-\ref{fig:kVsPtC5}. The qualitative behavior of the fluctuations in Au+Au collisions changes dramatically when comparing central to mid-central collisions. This behavior may be consistent with what is expected from jet suppression. In the most central collisions, jet suppression effectively dilutes the hard, non-Poissonian contribution to the distribution resulting in an increasing value of $k$. In non-central collisions, the dilution of the non-Poissonian hard component is reduced and $k$ decreases. This could also explain the rapid increase in $k$ as a function of centrality in the most central collisions. More detailed simulations are being studied to better understand the measured trends and to better estimate contributions from hard processes and geometry fluctuations.

\section{Summary}

PHENIX has measured charged particle multiplicity as a function of collision energy, centrality, collision species, and transverse momentum.  The scaled variance is seen to decrease when comparing 200 to 62 GeV Au+Au collisions.  However, the scaled variance increases when comparing 200 to 62 GeV Cu+Cu collisions. The scaled variance for 62 GeV Cu+Cu collisions behaves qualitatively like measurements at SPS energies.

\smallskip


\section{References}
\medskip
\begin{thebibliography}{9}
\bibitem{na49MF} Rybczynski M {\it et al} (NA49 Collaboration) 2004 {\it Preprint} nucl-ex/0409009.
\bibitem{phenixNIM} Adcox K {\it et al} (PHENIX Collaboration) 2003 {\it Nucl. Instrum. Meth.} A {\bf 499} 469.
\bibitem{ppg005} Adcox K {\it et al} (STAR Collaboration) 2002 {\it Phys. Rev.} C {\bf 66} 024901.
\bibitem{e802MF} Abbott T {\it et al} (E-802 Collaboration) 1995 {\it Phys. Rev.} C {\bf 52} 2663.
\end{thebibliography}
\end{document}